\documentclass[aps,twocolumn,floatfix,showpacs,superscriptaddress]{revtex4}
\usepackage{graphicx}
\usepackage{amsmath}
\usepackage{amssymb}
\usepackage{dsfont}
\usepackage{bm}
\usepackage{bbm}
\usepackage{color}

\DeclareMathOperator{\e}{e}

\newcommand{\bra}[1]{\langle #1|}
\newcommand{\ket}[1]{|#1\rangle}

\begin{document}

\title{Topological blockade and measurement of topological charge}
\author{B. van Heck}
\affiliation{Instituut-Lorentz, Universiteit Leiden, P.O. Box 9506, 2300 RA Leiden, The Netherlands}
\author{M. Burrello}
\affiliation{Instituut-Lorentz, Universiteit Leiden, P.O. Box 9506, 2300 RA Leiden, The Netherlands}
\author{A. Yacoby}
\affiliation{Department of Physics, Harvard University, Cambridge, Massachusetts 02138 USA}
\author{A. R. Akhmerov}
\affiliation{Instituut-Lorentz, Universiteit Leiden, P.O. Box 9506, 2300 RA Leiden, The Netherlands}

\begin{abstract}
The fractionally charged quasiparticles appearing in the 5/2 fractional quantum Hall plateau are predicted to have an
extra non-local degree of freedom, known as topological charge. We show how this topological charge can block the tunnelling of these
particles, and how such \emph{topological blockade} can be used to readout their topological charge. We argue that
the short time scale required for this measurement is favorable for the detection of the non-Abelian anyonic statistics of the
quasiparticles. We also show how topological blockade can be used to measure braiding statistics, and to couple a
topological qubit with a conventional one.
\end{abstract}


\maketitle

\textit{Introduction.-} The 5/2 fractional quantum Hall plateau is expected to be described by the Moore-Read wave function \cite{moore91} or its particle-hole conjugate Anti-Pfaffian state \cite{lee07,levin07}. This means that every pair of $e/4$ quasiparticles appearing in this phase have an extra neutral degree of freedom, topological charge, which does not affect local measurements and does not influence the energy of the system as long as they are well separated.

Topological charge manifests itself in the peculiar braiding statistics of these quasiparticles \cite{nayak96, bonderson11}: they are
non-Abelian anyons and their topological degree of freedom can be manipulated through ordered exchanges of
quasiparticles whose result is independent of the path used for braiding.
Such stability under local perturbations allows to exploit non-Abelian anyons to store and process quantum information
in a way that is highly protected from thermal noise and thus to potentially implement a topological quantum computer
\cite{kitaev03, nayak08}.

However there is no definite experimental proof that topological charge indeed exists.
Even when two quasiparticles are close to each other, there are no clear cut signatures of the topological charge: this
extra degree of freedom is completely charge-neutral, and hence very hard to detect.
The most actively developed tool predicted to readout the combined state of several non-Abelian anyons, and to prove they possess
fractional statistics, is non-Abelian Fabry-Perot interferometry \cite{nayak08,fradkin98,dassarma05,stern06,bonderson06,bishara09}. 
However, the currently existing non-Abelian interference experiments \cite{willett09,willett10,kang11,willett12} are not conclusive.
The interferometers are relatively sensitive to dephasing, since the length of the trajectory has to be
sufficiently large. Moreover they are described by a rather complicated theory \cite{fendley07} due to the
presence of many types of edge excitations \cite{gro11,ven12}.
In addition the interferometers are sensitive to all the anyons encircled by the interference loop, some of which may
even be coupled to edge states, further obscuring the interpretation of the results
\cite{overbosch07,bishara08,rosenow08,simon11,clarke11}. Other tools exist designed to probe macroscopic consequences of the existence of topological charge \cite{law06, feldman06, yang09, wang10}, however they do not allow to follow the behavior of a single anyonic excitation. Here we propose a setup for measuring the topological charge that does not suffer
from these limitations. Our setup is local, so it is only sensitive to the topological charge of two anyons, and it does not rely on using edge states.
Instead it is based on the phenomenon of topological blockade, explained below.

\begin{figure}[tb]
\includegraphics[width=\linewidth]{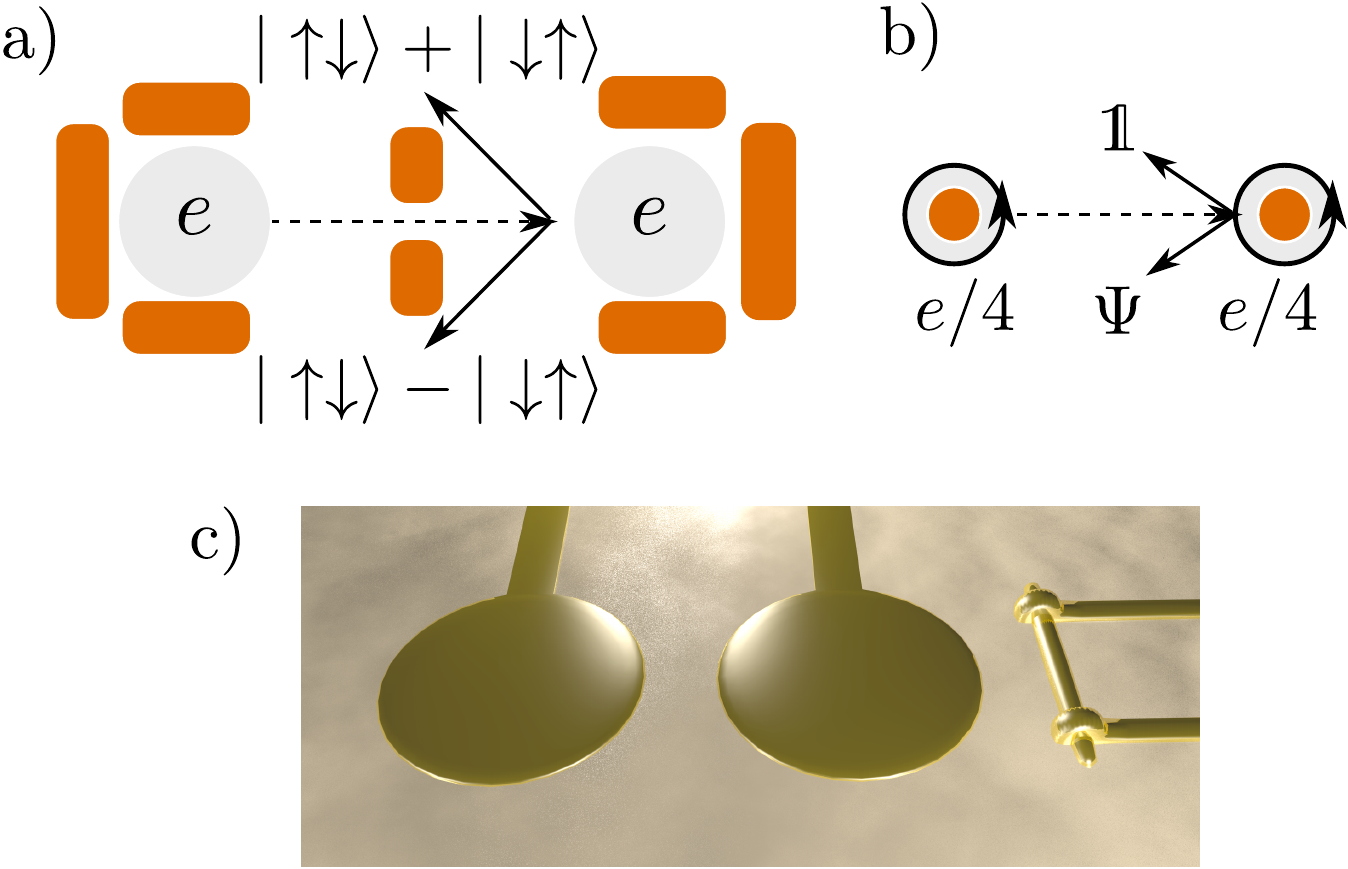}
\caption{\label{fig:comparison} Panel (a): two electrons with charge $e$ (grey circles) trapped by several gates
(rectangles) form a singlet-triplet qubit. The singlet and triplet states of the qubit acquire different energies when
one of the electrons tunnels. Panel (b): a topological qubit is formed by two quasiparticles of the Moore-Read quantum
Hall state with charge $e/4$. They are trapped by gates (filled circles). When one of these quasiparticles tunnels to
the other, two degenerate wave functions of the qubit corresponding to the vacuum and fermion fusion channels acquire
different energies. Panel (c): A sketch of a possible implementation of the topological blockade measurement setup featuring two local gates to form the quantum dots with size $\sim 100$ nm, and a charge sensor. The voltage applied to each dot is just enough to attract a single quasiparticle.}
\end{figure}

We begin our consideration from the simple observation that any inherent property of a particle that may impose an energy penalty, can
also prevent its motion. 
The most commonly known examples are the electric charge, which causes Coulomb blockade, and spin, resulting in spin blockade \cite{ono02,johnson05,koppens05,petta05,taylor07,foletti09,barthel09,bluhm10}. Less common examples include the position of a particle, causing elastic blockade \cite{koulakov98}.
Topological charge makes no exception: if the energy cost required to move two anyons onto the same region in space (fusing) is too high
due to their topological charge, then the anyons will not move. 
Since anyons have charge, detecting their position is not much harder than that of usual electrons, and standard
techniques such as QPC charge sensing \cite{field93,johnson05} or single electron transistor probes \cite{yoo97} can be used for this purpose \cite{ven11}. Blockade measurements are a standard technique in quantum systems, and they are much simpler than the measurement of a force acting on a single quasiparticle, proposed as an alternative to interferometry in Ref.~\cite{baraban09}.

The particular setup for the detection of topological charge that we propose is very similar to that of a singlet-triplet
spin qubit (see Fig.~\ref{fig:comparison}), where spin blockade is successfully used to distinguish a singlet state
of two electrons from a triplet one \cite{petta05,barthel09}. Two anyons are trapped to two dots formed by metallic gates \cite{dot_footnote}. The energies of the anyons are controlled by gate voltages, and the charge position is measured by a nearby charge sensor.

\begin{figure}[t]
\includegraphics[width=\linewidth]{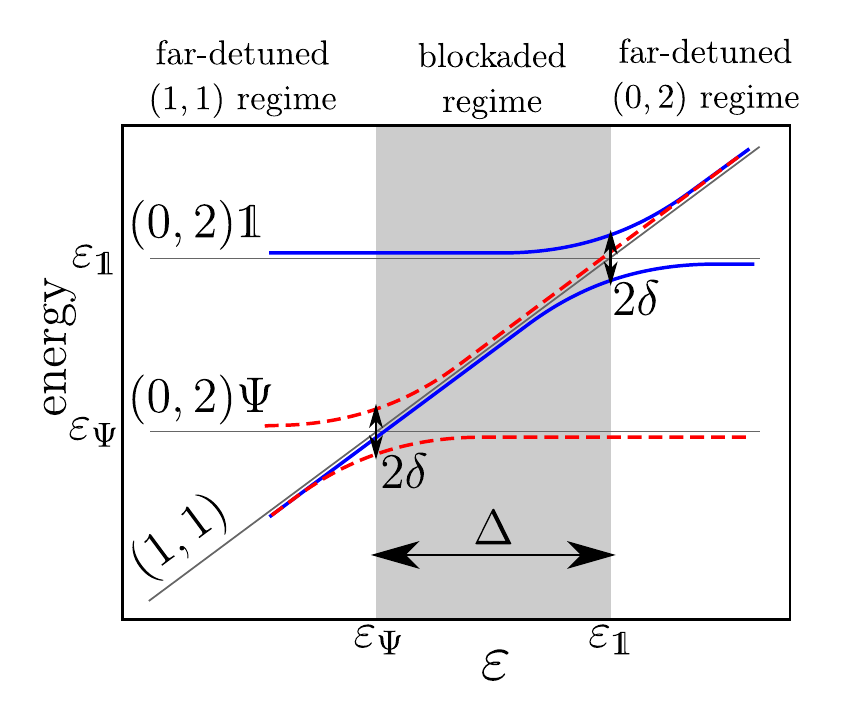}
\caption{\label{fig:levels} Main idea of topological blockade: spectrum of the Hamiltonian \eqref{ham} as $\varepsilon$, which states the energy of the (1,1) charge configuration, is varied. Two avoided crossings
occur when $\varepsilon$ is degenerate with the fusion energies $\varepsilon_\Psi, \varepsilon_\mathds{1}$ of the
anyons. Blue solid (red dashed) lines identify the eigenstates of the Hamiltonian with topological charge $\mathds{1}$ ($\Psi$). In the energy window between $\varepsilon_\Psi$ and $\varepsilon_\mathds{1}$ a blockaded regime occurs, with the (1,1) charge configuration favorable if the topological charge is $\mathds{1}$ but not if it is $\Psi$. For $\varepsilon\ll\varepsilon_\Psi$ and
$\varepsilon\gg\varepsilon_\mathds{1}$, the charge configurations $(1,1)$ and $(0,2)$ are respectively favored, independently on the topological charge shared by the anyons (far-detuned regimes).}
\end{figure}

In the following we analyze the performance of the proposed topological blockade readout of topological charge using a model
calculation. We continue by discussing experimental challenges and important energy scales for measuring topological
blockade. Finally, we propose several applications of topological blockade: a setup that should measure non-Abelian
braiding statistics, and a setup allowing to entangle a topological qubit with a singlet-triplet qubit.

\textit{The model.-} A topological qubit, shown in Fig.~\ref{fig:comparison}b, consists out of two quantum dots trapping a pair of quasiparticles with charge $e/4$ (Ising anyons). The energy levels of the dots can be separately controlled by varying gate voltages. When the gate voltage difference is small, the occupation number of both dots is equal, so that the system is in the $(1,1)$ configuration, where each index describes the occupation of each dot. When the voltage difference between the two dots is sufficiently large, a quasiparticle tunnels from the left dot to the neighboring one, and the ground state of the system becomes $(0, 2)$. Different states of the qubit are characterized by the fusion channel of two quasiparticles: vacuum ($\mathds{1}$) or fermion ($\Psi$). We consider a limited gate voltage interval such that the excited orbital states and the charge $(2, 0)$ arrangement are higher in energy than the four states relevant for the readout: $\left\lbrace \ket{(0,2)\mathds{1}}, \ket{(1,1)\mathds{1}}, \ket{(0,2)\Psi}, \ket{(1,1)\Psi} \right\rbrace$.

Similar to the singlet-triplet qubit case \cite{taylor07}, the Hamiltonian of the topological qubit is given by
\begin{equation} \label{ham}
 H=\begin{pmatrix} H_\mathds{1} & 0 \\ 0 & H_\Psi \end{pmatrix}, \quad {\rm with} \quad 
H_a=\begin{pmatrix} \varepsilon_a & \delta \\ \delta & \varepsilon \end{pmatrix}.
\end{equation}
Here $\varepsilon$ is the energy of states $\ket{(1,1)\mathds{1}}$ and $\ket{(1,1)\Psi}$, while $\varepsilon_\mathds{1}$ and $\varepsilon_\Psi$ are the energies of the states $\ket{(0,2)\mathds{1}}$ and $\ket{(0,2)\Psi}$ respectively. For definiteness we assume that $\varepsilon_\Psi <\varepsilon_\mathds{1}$ \cite{baraban09}, however our conclusions are not limited to this assumption. The tunneling between different charge configurations has amplitude $\delta$. As any local process, this tunneling preserves the topological charge.

The energy levels of the Hamiltonian \eqref{ham} are shown in Fig. \ref{fig:levels} as a function of $\varepsilon$,
which is controlled by gate voltages. The two charge configurations $(1,1)$, $(0,2)$ become degenerate in the $\Psi$
($\mathds{1}$) channel when $\varepsilon=\varepsilon_\Psi$ ($\varepsilon=\varepsilon_\mathds{1}$), and consequently
$\delta$ leads to avoided crossings in the spectrum. Between the two crossings, there exists an energy window of width
$\Delta=\varepsilon_\mathds{1}-\varepsilon_\Psi$ where the $(0,2)$ occupancy is favored with respect to the $(1,1)$
occupancy if the topological charge is $\Psi$ but not if it is $\mathds{1}$. This identifies the blockaded regime, where
charge tunnelling is allowed or blocked depending on the fusion channel of the anyons. The energy $\Delta$ is similar to
the singlet-triplet exchange splitting in spin blockade. This blocked region allows for efficient conversion from topological charge to real charge and hence allows readout of the topological state.

The topological charge of the double-dot system is subject to decoherence, due to coupling to the edges or other impurities in the quantum Hall liquid surrounding the system, which may cause transitions between the $\mathds{1}$ and $\Psi$ states in the same charge configuration. Assuming this process is independent of $\varepsilon$, we introduce a constant decay rate $\gamma$ and model the time evolution of our system using a Lindblad master equation 
\begin{equation}\label{masterequation}
\dot\rho=-i\left[H,\rho\right]+\tfrac{1}{2}\sum_j\,2L_j\rho L_j^\dagger - \{L_j^\dagger L_j, \rho\}
\end{equation}
with operators $L_1=L_2^\dag=\sqrt{\gamma}\ket{(1,1)\Psi}\bra{(1,1)\mathds{1}}$ and $L_3=L_4^\dag=\sqrt{\gamma}\ket{(0,2)\Psi}\bra{(0,2)\mathds{1}}$ describing the topological charge relaxation.

\begin{figure}[t]
 \includegraphics[width=\linewidth]{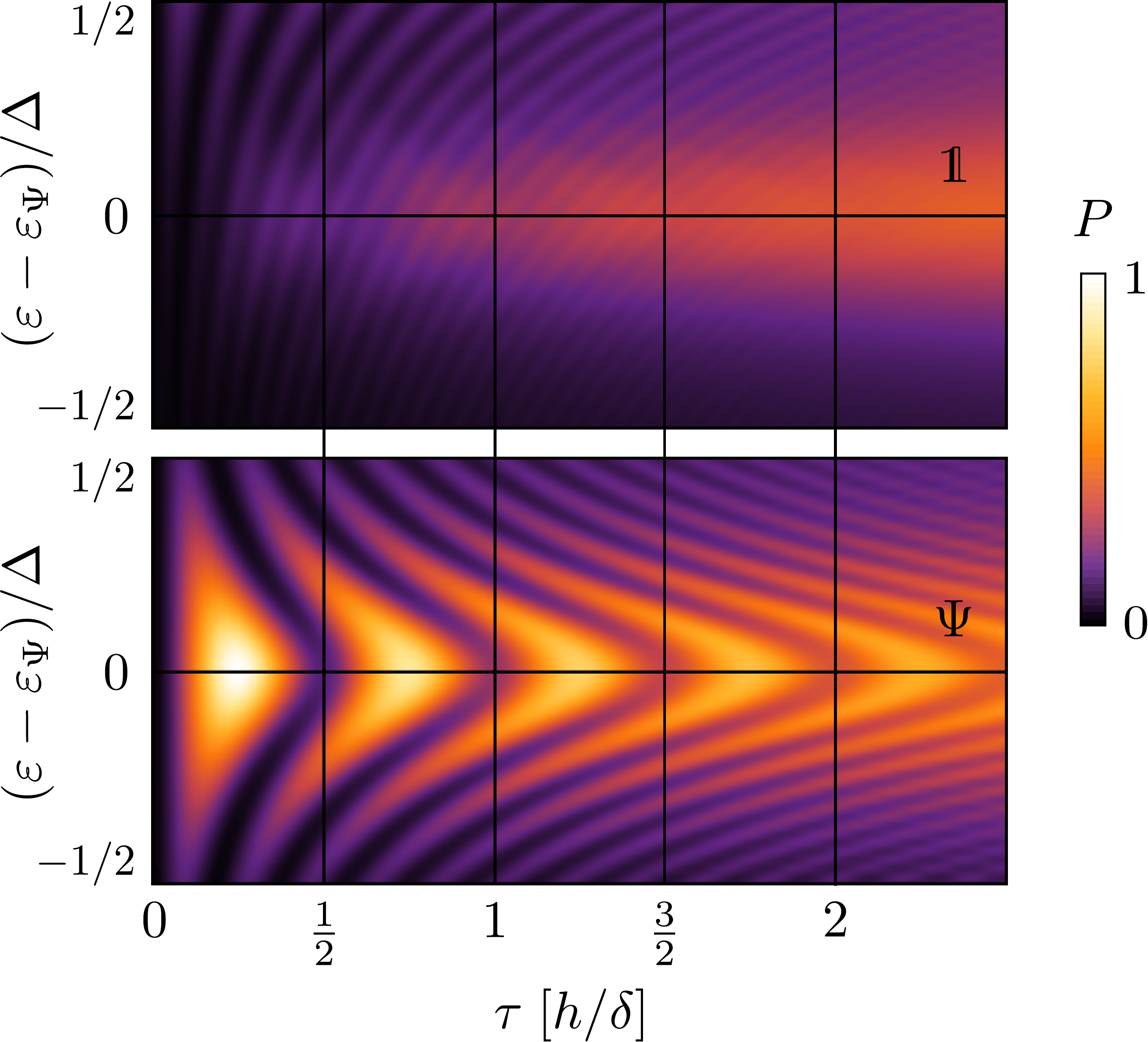}
\caption{\label{fig:time_evolution} Probability $P$ to measure the charge configuration $(0,2)$ when the system starts in the configuration $(1,1)$ with initial topological charge $\Psi$ (bottom panel) or $\mathds{1}$ (top panel), as a function of pulse duration $\tau$ and $\varepsilon$. Obtained from the numerical solution of the master equation \eqref{masterequation}, with parameters $\Delta=10 \delta$, $\gamma=0.1 \delta$.}
\end{figure}

\textit{Readout.-} The topological charge in the $(1,1)$ configuration at
$\varepsilon\ll\epsilon_\Psi$, can be determined by bringing it adiabatically through the avoided crossing into the blocked region, and
measuring the final charge configuration. This requires that the charge manipulation is performed on a time
scale $\tau \gg h/\delta$ (to avoid Landau-Zener transitions at the crossing).

If instead the topological charge relaxes too rapidly to perform the adiabatic passage, a faster readout procedure is needed. We describe here a method analogous to the rapid single-shot measurements of singlet-triplet qubits \cite{barthel09}. The system is initialized at $\varepsilon\ll\varepsilon_\Psi$ and the energy is then increased non-adiabatically to $\varepsilon\sim\varepsilon_\Psi$
for a short pulse of duration $\tau$, after which $\varepsilon$ is driven back to the initial value with a second fast
pulse. The topological charge is again inferred by a charge measurement of the final occupancies of the dots.

During the pulse, two anyons in the $\Psi$ channel oscillate between the $(1,1)$ and $(0,2)$ charge configurations with
period $h/2\delta$. If $\tau$ equals half of this period, the transition probability from $\ket{(1,1)\Psi}$ to
$\ket{(0,2)\Psi}$ is maximized. An analogous charge transition in the vacuum channel is strongly suppressed because the
state $\ket{(0,2)\mathds{1}}$ lies at a higher energy $\varepsilon_\mathds{1}$. In the ideal case, the initial
topological charge can be inferred by the final occupancies of the dot, with $(1,1)$ and $(0,2)$ corresponding to
$\mathds{1}$ and $\Psi$ respectively.  Unlike the adiabatic measurement, the time allowed for the charge measurement in this case is limited by the electric charge relaxation to the ground state, $(0,2)\to(1,1)$.

In a more realistic scenario, incoherent processes may alter the results and the position of the resonance cannot be
known in advance with great accuracy. Fig.~\ref{fig:time_evolution} shows the probability to measure the charge
configuration $(0,2)$ after a pulse of duration $\tau$ is performed at an energy $\varepsilon$, sweeping a range of
width $\Delta$ centered around $\epsilon_\Psi$, for the two different initial topological charges. Coherent oscillations
dominate in the $\Psi$ channel for $\tau\ll h/\gamma$, leading to fringes with peaks at $\tau^*_n=(n+1/2)h/2\delta$. The
brightest peak occurs at $\tau^*_0=h/4\delta$, making this the optimal duration of the pulse. If $\varepsilon$ is varied
for longer times and away from the resonance, the period of the fringes shortens and their intensity diminishes.
Since this readout method works identically if $\varepsilon\sim\varepsilon_{\mathds{1}}$ (only with
the roles of $\mathds{1}$ and $\Psi$ states interchanged), either can be used
to detect the topological charge of the prepared state.

The sum of the charge manipulation time $\tau$ and the charge readout time should be much shorter than the
topological charge relaxation time, which is equal to $h/\gamma$. The adiabatic charge manipulation requires $\tau \gg
h/\delta$, while the coherent manipulation requires a faster time scale $\tau \sim h/4\delta$, hence we arrive at the condition $\delta \gg \gamma$. If single shot readout is desired, the charge readout time should also be much shorter than the topological charge
relaxation time $h/\gamma$. However, a quick low
fidelity readout of the charge position is sufficient for the detection of the topological charge, since the measurements
can be repeated many times. Additionally, in order for for the two topological charges to be distinguishable, the blockaded region
should be larger than the region where charge tunneling occurs $\delta\ll\Delta$.

The appropriate parameter conditions can be reached by a careful design of the setup. It has been estimated that an effective potential in the two-dimensional electron gas with a width of a few magnetic lengths ($l_B$) can trap single quasiholes with a typical radius of $3l_B \approx 30$ nm \cite{wan08,morf11,bonderson11-2}. Under this assumption, numerical works calculated $\Delta\approx0.01e^2/\epsilon l_B$ with an upper bound of $1K$ \cite{baraban09,morf11}. For larger dots $\Delta$ is reduced since it is bounded from above by the level spacing.
The speed of relaxation of the topological charge $\gamma$ due to the coupling to disorder-induced anyons can be
estimated as $\Delta \e^{-l/\xi}$, with $\xi\approx 2.3l_B$ the characteristic length scale associated with the quantum Hall
liquid excitation gap \cite{baraban09} and $l$ the distance of the double-dot system from the nearest impurity.
Requiring $\gamma\approx 0.01 \Delta$ then yields a lower bound $l\approx100$ nm. Finally $\delta$ is
exponentially small in the distance between the dots, so the condition $\delta\ll\Delta$
requires the inter-dot distance to be larger than $\xi$.

These requirements are less stringent than the requirements for operation of a non-Abelian interferometer. There the
readout time must still be shorter than $h/\gamma$, however it should also be much larger than the time of flight of
a neutral excitation through the interferometer loop. This time of flight is given by $L/v \lesssim \hbar L / \Delta \xi$, with $L \gg \xi$ the length of the interferometer path. Additionally, $\gamma$ is increased due to the coupling
of the interferometer loop to bulk anyons \cite{rosenow08,clarke11}. The non-Abelian interferometers however have the advantage that they are able to measure the Abelian part of the braiding
statistics \cite{nayak08,kang11,simon11}, to which topological blockade is completely insensitive. The interferometers can also measure topological charge of more than two anyons, unlike the topological blockade.

Quasiparticles in the Abelian 331 state \cite{halperin83}, which is the most likely alternative to the Pfaffian state, have finite spin polarization, and hence may cause spin blockade. However due to the large Zeeman splitting, the equilibrium spin distribution is highly imbalanced, unlike the topological charge. This imbalance can be easily detected by performing a series of repeated blockade measurements.

Using existent technology, the smallest dots can be formed by local top gates with size $\sim 100$ nm. This is similar to the expected quasiparticle size. The effective confinement potential is expected to be still smoother than this scale because the 2DEG is located $\sim 50$ nm away from the gate. Nevertheless since the splitting is only suppressed linearly with the size of the ungapped region, we expect that this will not result in big suppression of the $\Delta$. As long as the gate potentials are sufficiently small, these local gates just attract excess quasiparticles without forming the edge states. A local charge sensor similar to the one used in Ref.~\cite{ven11} could be fabricated in proximity to one of the dots, as shown in Fig.~\ref{fig:comparison}c.

\begin{figure}[t!]
 \includegraphics[width=\linewidth]{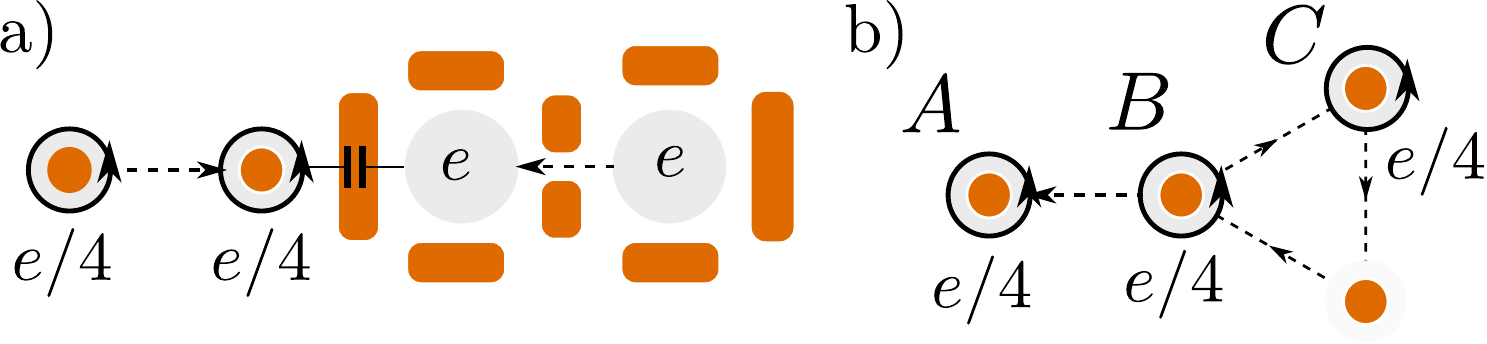}
\caption{\label{fig:other_setups} Applications of topological blockade. Panel (a): two-qubit system formed out of a topological and a spin qubit. For both qubits the computational degrees of freedom correspond to different charge configurations. Entanglement between the qubits can be induced by a capacitive coupling between the two double-dots. Panel (b): setup for the detection of non-Abelian statistics of the $\nu=5/2$ fractional excitations. Three anyons ($A, B, C$) are hosted in four dots and can be moved by varying gate potentials. Two counterclockwise exchanges of $B$ and $C$, implemented using the fourth empty dot, act as a NOT gate on the qubit formed by $A$ and $B$ \cite{dassarma05}.}
\end{figure}

\textit{Detection of non-Abelian statistics.-} In order to detect non-Abelian braiding statistics of the anyons, one
needs to combine the topological blockade-based readout device with a minimal setup for exchanging two anyons
\cite{fre06}. This setup is shown in Fig.~\ref{fig:other_setups}a, and it consists of a topological qubit with two
extra dots hosting a single $e/4$ quasiparticle. The quasiparticles are moved by varying the potential of the dots.
Both braiding and detection can then be performed in the four dot setup, using the same 
protocol proposed in \cite{dassarma05} in the context of interferometric devices.

\textit{Coupling with conventional qubits.-} Since topological blockade allows to translate the topological
charge into the position of the electric charge, it becomes possible to couple a topological qubit with conventional quantum
systems, similarly to what was done with Majorana qubits in superconducting systems \cite{jiang11,bondersonlutchyn11,hassler11}. 
Both topological and spin blockade translate the qubit degree of freedom into an electric charge configuration.
In the case of singlet-triplet qubits, this effect has been used to couple two neighboring double dots in order to
produce two-qubit entanglement \cite{yacoby12}. The same method can be explored to couple capacitively a topological
qubit to a singlet-triplet qubit hosted in a nearby double quantum dot (see Fig.~\ref{fig:other_setups}b). Independent
measurements on the two qubits can still be performed via two charge-sensing quantum point contacts. Additionally, the oscillatory motion of electric charge at the transition
between $(1,1)$ and $(0,2)$ states can also be used to couple the topological charge to electromagnetic radiation, thus
allowing coupling of a topological qubit with cavity qubits. Since the gate pattern needed to define the double-dot hosting the singlet-triplet qubit will likely introduce undesired edges in the quantum
Hall liquid, it would be necessary to have the second qubit in a different layer of the nano-structure.
Another difficulty to overcome is the presence of a strong magnetic field which increases the Zeeman
splitting of the triplet states and makes it comparable with the exchange splitting in the singlet-triplet system,
potentially ruining the operation of the spin qubit.

\textit{Conclusions.-} In conclusion, we have showed how to use topological blockade to measure topological charge.
While we focused on the most experimentally relevant case of the 5/2 fractional quantum Hall plateau, the same method
applies to any non-Abelian phase as long as the anyons also have electric charge. We have shown that the topological blockade is more robust than the non-Abelian interferometry, in part due to being insensitive to the Aharonov-Bohm phase. The downside is that it cannot probe the Abelian part of the braiding statistics. Finally, we have also shown how
to use topological blockade to measure braiding statistics and to couple topological qubits with a singlet-triplet spin
qubit. 

\textit{Acknowledgements.-} We have benefitted from discussions with C. W. J. Beenakker, C. Marcus, and B. Halperin.
This project was supported by the Dutch Science Foundation NWO/FOM and the ERC Advanced Investigator Grant. A.Y. acknowledges support from Microsoft Station Q.


\begin{thebibliography}{25}
\bibitem{dot_footnote}{The charge of the quasiparticle attracted to the gate voltage does not
matter for our considerations: both dots, which trap $e/4$ quasiparticles and antidots trapping $e/4$ quasiholes work
identically.}
\end{thebibliography}
\end{document}